\documentclass{emulateapj}

\newcommand{\be}{\begin{eqnarray}}
\newcommand{\ee}{\end{eqnarray}}
\newcommand{\beq}{\begin{equation}}
\newcommand{\eeq}{\end{equation}}
\def\simless{\mathbin{\lower 3pt\hbox
      {$\rlap{\raise 5pt\hbox{$\char'074$}}\mathchar"7218$}}}
\def\simgreat{\mathbin{\lower 3pt\hbox
      {$\rlap{\raise 5pt\hbox{$\char'076$}}\mathchar"7218$}}} %> or of order

\begin{document}

\title{ Massive Satellites of Close-In Gas Giant Exoplanets }

\author{Timothy A. Cassidy\altaffilmark{1}, Rolando Mendez\altaffilmark{2}, 
Phil Arras\altaffilmark{2}, Robert E. Johnson\altaffilmark{1,2}, Michael F.
Skrutskie\altaffilmark{2}}

\altaffiltext{1}{Engineering Physics Program, University of Virginia \\
Charlottesville, VA 22904-4325}
\altaffiltext{2}{Department of Astronomy, University of Virginia \\
P.O. Box 400325, Charlottesville, VA 22904-4325}
%\email{tac2z@virginia.edu, rem5d@cms.mail.virginia.edu, 
%arras@virginia.edu, rej@virginia.edu, mfs4n@virginia.edu}

\begin{abstract}

We study the orbits, tidal heating and mass loss from satellites around
close-in gas giant exoplanets. The focus is on large satellites which are
potentially observable by their transit signature. We argue that even
Earth-size satellites around hot Jupiters may be immune to destruction
by orbital decay; detection of such a massive satellite would strongly
constrain theories of tidal dissipation in gas giants, in a manner
complementary to orbital circularization. The star's gravity induces
significant periodic eccentricity in the satellite's orbit.  The resulting
tidal heating rates, per unit mass, are far in excess of Io's and dominate
radioactive heating out to planet orbital periods of months for reasonable
satellite tidal $Q$. Inside planet orbital periods of about a week,
tidal heating can completely melt the satellite. Lastly, we compute an
upper limit to the satellite mass loss rate due to thermal evaporation
from the surface, valid if the satellite's atmosphere is thin and vapor
pressure is negligible.  Using this upper limit, we find that although
rocky satellites around hot Jupiters with orbital periods less than a
few days can be significantly evaporated in their lifetimes, detectable
satellites suffer negligible mass loss at longer orbital periods.

\end{abstract}


% -----------------------------------------------------------

\section{Introduction}

The high photometric precision of the Hubble Space Telescope allows
interesting constraints to be placed on the existence of massive
satellites orbiting transiting exoplanets. Two systems have been
searched so far, HD 209458b \citep{2001ApJ...552..699B} and HD 189733b
\citep{2007A&A...476.1347P}. From the transit lightcurve and timing,
upper limits of roughly an Earth radius and mass have been placed on
hypothetical satellites around these two planets. A number of studies
have discussed detection techniques \citep{1999A&AS..134..553S,
2002ApJ...580..490H, 2004ApJ...616.1193B, 2004IAUS..213...80D,
2006A&A...450..395S, 2007A&A...470..727S,2009MNRAS.392..181K,2009MNRAS.396.1797K}. 
This paper is concerned with
the orbits and physical structure of these hypothetical moons.

With the launch of the Kepler satellite \citep{2004ESASP.538..177B},
$\sim 10^5$ stars will be monitored with photometric precision sufficient
to detect Earth size objects. The estimated hundreds of hot Jupiters
(gas giant planets with orbital periods less than 1 week) to be found
by Kepler will provide a large sample to investigate the existence of
Earth-size satellites \citep{2006A&A...450..395S}.

\citet{2006PASP..118.1136J} note another possible observational
manifestation of satellites. Interaction between a satellite and
the planet's magnetosphere may give rise to an extended gaseous
torus orbiting the planet, as in the Jupiter-Io system. Material
stripped from the satellite's surface and atmosphere would contribute
to the transmission spectrum of the planet. It has been suggested
\citep{2003Natur.422..143V} that absorption in the upper atmosphere
of HD 209458b \citep{2002ApJ...568..377C, 2003Natur.422..143V,
2004ApJ...604L..69V, 2008ApJ...673L..87R, 2007Natur.445..511B} is due
to a high rate of atmospheric escape which increases the density at
large radii.  The plasma torus model is an alternative explanation for
such absorption high in the atmosphere.

There has been interest recently in the properties of satellites
around exoplanets, mainly focused on the habitability of
moons (e.g. \citealt{1987AdSpR...7..125R, 1997Natur.385..234W,
2006ApJ...648.1196S}) around planets at large orbital separation from
the star. Our focus in this paper is on planets to be found with Kepler,
hence we restrict our attention on close-in planets for which the transit
probability is substantially higher.  As we will show, moons in such
systems will be subjected to intense tidal heating, with consequences
for the moon's structure, evaporation rate, and interaction with its
surroundings.

Two recent studies of the orbital stability of satellites are important
in this work.  \citet{2006MNRAS.373.1227D} investigated orbital stability
for point mass star, planet and satellite, for a range of eccentricity of
both the planet and satellite. They point out that, for small planetary
eccentricity, satellite orbits are stable for $a_s \la 0.5 a_H$, where $a_s$
is the semi-major axis of the satellite around the planet and $a_H$ is
the planet's Hill radius. 
%Another way of viewing this result is that the eccentricity of the 
%moon's orbit is order unity as the stability
%limit is approached. 
As we will show, the satellite orbits have large periodic eccentricity
induced by the star as the stability limit $a_s \simeq 0.5 a_H$
is approached.
Hence, while Io's eccentricity and tidal heating
are due to interaction with Jupiter and the other Galilean satellites,
moons of close in exoplanets have a large periodic eccentricity due to
stellar forcing. We show that this periodic eccentricity leads to large
tidal heating for systems close to the parent star. This tidal heating
may have a significant impact on the structure of the satellite, 
inducing melting of the interior and perhaps reinforcing dynamo generation
of magnetic fields in the core.

As pointed out by \citet{2002ApJ...575.1087B}, such close-in moons,
orbiting slowly rotating planets synchronized to the star, are subject
to orbital decay. The tide raised on the planet by the moon acts to spin
up the planet while shrinking the moons orbit, eventually leading to the
moon impacting the planet. The orbital decay rate is sensitive to the
tidal $Q_p$ of the gas giant planet. \citet{2002ApJ...575.1087B} used
$Q_p=10^5$, the estimate for the Jupiter-Io tidal interaction, leading
to the constraint that hot Jupiters cannot have moons larger than $\sim
10^{-4}M_\oplus$, where $M_\oplus$ is the Earth's mass. Such tiny moons
would be undetectable by the transit method.  However, as we discuss,
the appropriate value of $Q_p$ may be larger by a factor $\sim 10^8$,
allowing stable orbits for potentially detectable moons with radii $\ga
R_\oplus$.

Little attention has been given to the formation scenario of the massive
satellites discussed in this paper. \citet{2006Natur.441..834C} note that
the outer planets of the Solar System contain only a fraction $10^{-4}$
of the planet's mass in the satellite systems. If applied to extrasolar
planets, this phenomenological scaling implies an upper limit for the
satellite mass of $\la
M_\oplus/30$ which would be undetectable by the transit method. An
alternative possibility is for the gas giant to capture a terrestrial
planet into a bound orbit.

%on the mechanisms to form
%{\bf Background}. Gas giant planet must migrate in. Fate of initial moons?
%Acquire moons during migration? Migration must be early to keep planets
%inflated, so moons may not have time to significantly cool, still molten
%when the park near the star. We compute heating if solid, but likely
%far lower rate since they are molten.

Satellite orbits in the restricted 3 body problem are discussed
in section \ref{sec:orbits}. Orbital decay due to the tide raised in the
planet by the satellite is discussed in section
\ref{sec:decay}. Tidal heating of the satellite due to the non-circular
orbit and its consequences for internal structure
are discussed in section \ref{sec:heating}.  Coupled tidal heating and
satellite orbital evolution are described in section \ref{sec:energetics}.
The rate of evaporative mass loss from the satellite is discussed in section \ref{sec:massloss} and
conclusions are stated in section \ref{sec:conclusions}.

% -----------------------------------------------------------

\section{ Satellite orbits}
\label{sec:orbits}

In this section we review the moon's orbit, treating moon, planet and
star as point masses. This problem is referred to as the ``main problem
in lunar theory" \citep{1961mcm..book.....B}, and considerable analytic
progress can be made. We will give a simple derivation of the satellite's
orbit, as needed to compute tidal heating, for the case where the planet's
orbit around the star is circular (the ``circular restricted 3-body problem")
and the ``unperturbed" orbit of the satellite is circular. That is, we will
compute the small deviations from a circular orbit due to the gravity of the
star.

We consider star, planet and satellite of masses $M_* \gg M_p \gg M_s$,
respectively. We will ignore the influence of the satellite on the
star-planet orbit, treating it as Keplerian with semi-major axis $a_p$
and eccentricity $e_p$. The mean angular motion of the planet's orbit is
$n_p \simeq (GM_*/a_p^3)^{1/2}=2\pi/P_p$, where $P_p$ is the planet's orbital period.
The satellite's orbit around
the planet can be treated as nearly Keplerian, with semi-major axis $a_s$,
eccentricity $e_s$ and mean motion $n_s \simeq (GM_p/a_s^3)^{1/2}=2\pi/P_s$.

We work in a reference frame with origin at the center of the planet
and non-rotating axes fixed with respect to distant observers.
In this reference frame the star only contributes a tidal force.
Cylindrical coordinates $(r_s,\phi_s)$ and $(r_*,\phi_*)$ are used for the 
positions of the satellite and star, respectively.
The orbits are assumed coplanar for simplicity. The
equations of motion for the satellite are then
\be
\ddot{r}_s & = & \frac{\ell_s^2}{r_s^3} - \frac{GM_p}{r_s^2} + \frac{\partial {\cal R}}{\partial r_s}
\label{eq:ddotr}
\\
\dot{\phi_s} & = & \frac{\ell_s}{r_s^2}
\label{eq:dotphi}
\\
\dot{\ell_s} & = &  \frac{\partial {\cal R}}{\partial \phi_s}
\label{eq:dotell}
\ee
where $\ell_s$ is the satellite's orbital angular momentum
per unit mass and ${\cal R}={\cal R}(r_s,\phi_s,r_*,\phi_*)$ is the negative of the
tidal gravitational potential from the star. We can simplify the form
of ${\cal R}$ using the fact that $r_s \ll r_*$ to find
the leading order result in $r_s/r_*$
\citep{1961mcm..book.....B},
\be
{\cal R} & = & \frac{GM_*r_s^2}{2r_*^3} \left( 3 \cos^2[\phi_s-\phi_*] - 1 \right).
\label{eq:R}
\ee
Given an orbit for the planet-star system, $r_*(t)$ and $\phi_*(t)$, eq.\ref{eq:ddotr},
\ref{eq:dotphi} and \ref{eq:dotell} can be integrated in time to find the orbit
of the satellite.
%Plugging in Keplerian orbits for star and moon into the disturbing
%function, we can derive the variation of the orbital elements
%\citep{1961mcm..book.....B}.

The parameter space for stable satellite orbits is strongly constrained.
For close-in planets, the planetary radius $R_p$ can be a large fraction
of the planet's Hill radius, $a_H=a_p(M_p/3M_*)^{1/3}$, implying that
all of its bound
satellite orbits are significantly perturbed by the star's gravity.
\citet{2006MNRAS.373.1227D} found stable orbits for $a_s \la 0.49
a_H (1.0- 1.0e_p - 0.27e_s )$. In terms of the orbital periods,
the stability limit is $P_s \la P_p/5$ for $e_p,e_s \ll 1$. The
dimensionless strength of the disturbing potential is then 
${\cal R}/(GM_p/a_s) \sim (n_p/n_s)^2 =
(P_s/P_p)^2 \leq 1/25$.  Ignoring the cohesive strength, the moon must
also orbit outside the planet's Roche radius at $a_{\rm Roche}=3^{1/3}
R_p (M_pR_s^3/M_sR_p^3)^{1/3}$, where $R_p$ is the radius of the planet.
For gas giants with mean density $\rho_p = 3M_p/(4\pi R_p^3) \sim 1\
{\rm g\ cm^{-3}}$ and terrestrial satellites with mean density $\rho_s
=3M_s/(4\pi R_s^3) \sim 5\ {\rm g\ cm^{-3}}$, the Roche radius is at or
below the planets surface. Hence we can consider the satellite disrupted
if it hits the planet's surface.  In summary, satellite orbits are bounded
at short orbital periods by the planet's surface and at long orbital
periods by the stability constraint, giving the range of possible orbital
periods $2\pi (R_p^3/GM_p)^{1/2} \la P_s \la P_p/5$, for $e_s,e_p \ll 1$.

The derivation in \citet{1961mcm..book.....B} uses the disturbing function
formalism and computes the variation of the orbital elements $a_s$, $e_s$, etc rather
than $r_s$ and $\phi_s$ directly. Here we
give a simpler derivation of the variation involving $r_s$ and $\phi_s$.
We consider a circular planetary orbit with $r_*=a_p$ and
$\phi_*=n_pt$, and a satellite orbit described by a circular orbit plus
small perturbations induced by the star. In detail, we write
\be
r_s(t) &  = & a_s + \delta r_s(t)
\label{eq:rs}
\\
\phi_s(t) & = & \phi_{s0} +  n_s t + \delta \phi_s(t)
\label{eq:phis}
\\
\ell_s(t) & = & a_s^2 n_s + \delta \ell_s(t).
\label{eq:ells}
\ee
Here $\phi_{s0}$ is a constant phase.
Plugging eq.\ref{eq:rs}, \ref{eq:phis} and \ref{eq:ells} into 
eq.\ref{eq:ddotr},
\ref{eq:dotphi} and \ref{eq:dotell}, the equations are satisfied 
at leading order. Eq.\ref{eq:dotell} can be immediately integrated
to yield
\be
\frac{\delta \ell_s(t)}{\ell_s} & = & \frac{3}{4} \left( \frac{n_p}{n_s} \right)^2
\cos\left[ 2(n_s-n_p)t  + 2 \phi_{s0} \right],
\label{eq:deltaell}
\ee
where we have approximated $n_p \ll n_s$ to simplify the expression.
Plugging eq.\ref{eq:deltaell} into eq.\ref{eq:ddotr}, we find the following
forced harmonic oscillator equation for $\delta r$,
\be
\ddot{\delta r_s} + n_s^2 \delta r_s & = & n_p^2 a_s \left( 
\frac{1}{2} + 3 \cos\left[ 2(n_s-n_p)t + 2\phi_{s0} \right] \right).
\label{eq:ddotdeltar}
\ee
The first term on the right hand side of eq.\ref{eq:ddotdeltar} gives
a shift in the orbital radius $\delta r_s = (1/2) a_s (n_p/n_s)^2$.
%, as if the 
%planet's mass was slightly smaller.
This constant term causes no tidal heating, so we ignore it from
here on. The second term on the right hand side of eq.\ref{eq:ddotdeltar}
causes an oscillatory change in $\delta r_s$. We can solve for this
term by writing $\delta r_s(t) = \delta r_{s0}\cos\left( \left[ 2(n_s-n_p)t +2\phi_{s0}
\right] \right)$, where $\delta r_{s0}$ is a constant.
The cosine factor then cancels out of the equation and the amplitude
$\delta r_{s0}$ can be found. The result is
\be
\frac{\delta r_s(t)}{a_s} &  = & - \left( \frac{n_p}{n_s} \right)^2 
\cos \left[ 2(n_s-n_p)t + 2\phi_{s0} \right].
\label{eq:deltar}
\ee
Plugging eq.\ref{eq:deltaell} and \ref{eq:deltar} into eq.\ref{eq:dotphi}, we find the angle
\be
\delta \phi_s(t) & = & \frac{11}{8} \left( \frac{n_p}{n_s} \right)^2
\sin \left[ 2(n_s-n_p)t + 2\phi_{s0} \right] .
\label{eq:deltaphi}
\ee
Eq.\ref{eq:deltar} and \ref{eq:deltaphi} confirm that the forced perturbations
vary as $(P_s/P_p)^2$ for small eccentricity orbits. These
formulas are valid when $P_s \ll P_p$, and underestimate the perturbations
as the orbit approaches instability.  We will use eq.\ref{eq:deltar}
and \ref{eq:deltaphi} to compute the tidal heating rate in section
\ref{sec:heating}.

\begin{figure}
\plotone{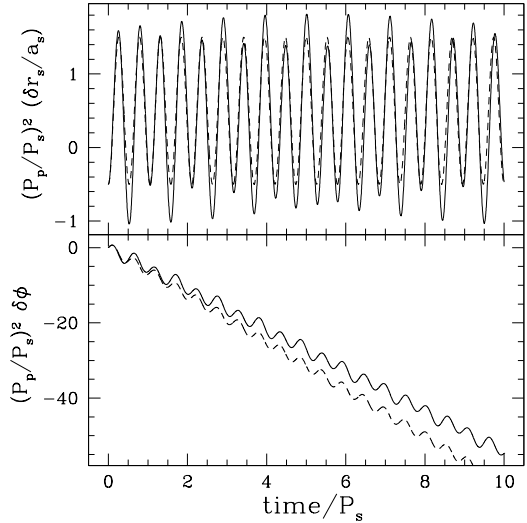} 
\caption{ Perturbations to the radius (upper panel) and longitude (lower panel) due to the
stellar gravitational field. Here $\delta r_s=r_s(t)-a_s$ and $\delta \phi_s=\phi_s(t)-n_s t$.
The solid lines are the result of numerical integration of eq.\ref{eq:ddotr},
\ref{eq:dotphi} and \ref{eq:dotell} for $P_p/P_s=20$, scaled by the factor $(P_p/P_s)^2$. 
The dashed lines are the analytic results including the periodic components
from eq.\ref{eq:deltar} and \ref{eq:deltaphi}, as well as the time-independent shift to the radius
$\delta r_s/a_s = (n_p/n_s)^2/2$, and the resultant change in orbital period. This change in orbital
period is the cause of the linear behavior of $\delta \phi$ in the lower plot, while the oscillatory
term from eq.\ref{eq:deltaphi} causes the short period variations.}
\label{fig:orbit}
\end{figure} 

Figure \ref{fig:orbit} compares a direct numerical integration of eq. \ref{eq:ddotr},
\ref{eq:dotphi} and \ref{eq:dotell} against the analytic formulas in eq.\ref{eq:deltar}
and \ref{eq:deltaphi}, showing that they agree to the stated accuracy in the 
small parameter $P_s/P_p$.

The discussion so far has examined small perturbations about circular
planet and satellite orbits. The same method can be used to find small
perturbations, induced by the star, about orbits with finite ``background"
eccentricities $e_{s0}$ or $e_{p0}$.  A detailed derivation and results
can be found in \citet{1961mcm..book.....B}.  Again orbital variations
of size $(P_s/P_p)^2$ are found due to stellar perturbations.  Tidal
heating of the satellite can arise due to either background $e_{s0}$
and $e_{p0}$, or perturbations induced by the star,  which exists
even for $e_{s0}=e_{p0}=0$. We show in section \ref{sec:heating} that
the relative size of these two
effects scales as $e_{s0}$ or $e_{p0}$ versus $(P_s/P_p)^2$. The forced
variation should provide a {\it lower limit} to the tidal heating,
and may significantly underestimate the heating for finite eccentricity
orbits at large $a_p$.

% -----------------------------------------------------------

\section{ Orbital decay due to tides raised in the gas giant planet}
\label{sec:decay}

Tidal friction is recognized to be an important factor in the survival of
hypothetical early satellite systems for inner Solar System
planets Mercury and Venus (e.g., \citealt{1973Natur.242...23B,
1973MNRAS.164...21W}). These studies pointed out that tides raised
in the planet by the satellite and the Sun can cause significant changes
in the satellite's orbit. The problem is even more severe as the planet
is moved closer to the star, as the stellar tides are stronger and the
Hill sphere is closer to the planet. However, a key difference occurs
if the planet in question is a gas giant rather than a rocky body. Gas
giants are less dissipative than rocky bodies, and hence satellites are
relatively more immune to destruction by orbital decay into the planet.

\citet{2002ApJ...575.1087B} studied orbital decay of satellites around hot
Jupiters. They point out that if the gas giant planet is synchronized to
the parent star, then the satellite orbits faster than the planet spins.
Tides raised in the planet by the satellite then attempt to spin up the
planet while shrinking the orbit of the satellite to conserve angular
momentum. The conclusion of their study was that only satellites
with masses $M_s<<M_\oplus$ could survive being dragged down to the
planet. Such tiny satellites would be undetectable by the transit method.
We point out that recent theoretical work on the tidal $Q_p$ in gas giants
implies that orbital decay is far less effective than the estimate by
\citet{2002ApJ...575.1087B} for the forcing periods of interest,
and hence orbital decay may be ineffective even for Earth mass satellites.

Using the equations from \citet{1966Icar....5..375G}, orbital decay 
over a timescale $T$ will lead to destruction of satellites with 
mass
\be
M_s & \geq & M_{s, \rm crit} = 5.4\times 10^{-3}\ M_p\ \left( \frac{Q_pP_s^{13/3}}{P_{\rm
dyn}^{10/3} T} \right),
\ee
where $Q_p$ is the quality factor for the gas giant planet and $P_{\rm dyn}=2\pi
(R_p^3/GM_p)^{1/2}$ is the dynamical time of the planet ($2.8\ {\rm
hr}$ for Jupiter). \citet{2002ApJ...575.1087B} evaluated this expression using
$Q_p \simeq 10^5$, the value inferred for orbital expansion in the Jupiter-Io
system \citep{1966Icar....5..375G}, as well as circularization of the 
extrasolar planets (e.g., \citealt{2003ASPC..294..213W}), finding
\be
M_{s,\rm crit} & = & 5\times 10^{-5} M_\oplus \left( \frac{M_p}{M_{\rm Jup}}
\right)
\left( \frac{Q_p}{10^5} \right) 
\nonumber \\ & \times & \left( \frac{P_s}{P_p/5} \frac{P_p}{4\ {\rm days}} \right)^{13/3}
\left( \frac{2.8\ {\rm hr}}{P_{\rm dyn}} \right)^{10/3}
\left( \frac{5\ {\rm Gyr}}{T} \right),
\label{eq:Mscrit}
\ee
where $M_{\rm Jup}$ is the mass of Jupiter, and we have scaled $P_s$ in
terms of the maximum orbital period $P_p/5$. If true, this result would
imply that Earth mass satellites would only be stable outside orbital
periods of $\sim 40$ days, with corresponding smaller transit probability
than for the hot Jupiters at $P_p \sim $ a few days.

The physical origin of the tidal $Q_p$ in gas giant planets has been an
outstanding question since at least the 1970's. The $Q_p$ for Jupiter is
constrained to be $Q_p \sim 10^5-10^6$ in order that Io's orbit expanded
into the Laplace resonance \citep{1966Icar....5..375G}. Early theoretical
work by \citet{1974Icar...23...42H} showed that the turbulent viscosity
generated by convective eddies required for the outward transport of
heat would give rise to $Q_p \sim 10^5$, perhaps explaining the observed
value. \citet{1977Icar...30..301G} then pointed out that turbulent eddies
in the planet have long turnover times, $t_{\rm eddy}$, compared to the
forcing periods of interest, severely decreasing the turbulent viscosity
used in Hubbard's calculation. \citet{1977Icar...30..301G} estimated $Q_p
\sim 10^{13}$ for ``equilibrium-tide" flow in Jupiter, underpredicting
the observed tidal dissipation rate by a factor of $10^7-10^8$. 
\citet{2005ApJ...635..688W} revisited Goldreich and Nicholson's calculation,
revising $Q_p$ downward to $Q_p \simeq 10^{12}$.

The detailed physics underlying tidal dissipation in gas giants may
strongly affect predictions of the orbital stability of the satellites
studied in this paper. Recent calculations by \citet{2004ApJ...610..477O}
and \citet{2005ApJ...635..688W} have found that when the tidal forcing
frequencies are resonant with inertial waves (waves with frequencies less
than twice the spin frequency, restored by the Coriolis force) that $Q_p
\sim 10^5-10^9$ can be obtained, due to the higher dissipation rate for
short lengthscale inertial waves. Conversely, when the forcing frequency
is not in the inertial range, no such resonant excitation occurs and $Q_p$
will be orders of magnitude larger, comparable to the equilibrium
tide value $Q_p \sim 10^{12}$, or perhaps somewhat smaller
if atmospheric waves can dissipate efficiently. The work of Ogilvie and
Lin and Wu has demonstrated that resonant excitation of inertial waves
is a promising mechanism to explain the small observed $Q_p$ values. A
generic feature in their results is that the tidal dissipation rate
is strongly dependent on the forcing frequency. Hence the practice of
taking a measured $Q_p$ from one situation and applying it to another
may be incorrect if the forcing frequencies are very different. The
consequences of the tidal dissipation factor varying with forcing
frequency have already been explored in the context of solar type stars
by \citet{2007ApJ...661.1180O}.

We now explore the consequences for survival of satellites of short period
exoplanets {\it if} the inertial wave hypothesis is correct. 
%In that
%case, the small $Q_p$ values do not apply to tides raised in a slowly
%rotating planet by the satellite. 
If synchronized, the spin period of the
gas giant equals its orbital period, $P_p$. But the orbital period of the
satellite must satisfy $P_s \leq P_p/5$ for orbital stability. For a quadrupole tide
raised in the planet, the forcing period would be shorter than $P_p/10$,
while inertial waves only exist with periods longer than $P_p/2$. Hence
inertial waves cannot be excited in synchronous planets by the satellite
and $Q_p \sim 10^5$ does not apply. 
%The forcing frequency is also far too
%low to be resonant with the f-mode (periods $\sim $ a few hours), hence
%no normal modes of oscillation are available for resonance with the tide
%over the bulk of the planet.

%The estimate of \citet{1977Icar...30..301G} found that dissipation
%due to turbulent viscosity was concentrated just below the tropopause,
%near $\sim 1\ {\rm bar}$ for Jupiter. For the hot Jupiters, strong
%insolation pushes the tropopause down to $\sim$kbar pressures (e.g.,
%\citealt{2006ApJ...650..394A}), where $t_{\rm eddy}$ is much longer.
%Using Goldreich
%and Nicholson's estimate of $Q=10^{13}$.
%, keeping in mind that it may
%underestimate $Q$ (overestimate dissipation). 
%Plugging this value into eq.\ref{eq:Mscrit} we find
%we find

The existence of exomoons around close-in gas giant exoplanets with
$P_p \la $ a couple months
provides a test of tidal dissipation theory.
From eq.\ref{eq:Mscrit}, in
order for a satellite of mass $M_s$ to survive for a time $T$ requires
\be
Q_p \geq Q_{\rm p, crit}
& = & 
2 \times 10^9\
\left( \frac{M_s}{M_\oplus} \right)
\left( \frac{M_{\rm Jup}}{M_p} \right)
\left( \frac{P_p}{5P_s} \frac{4\ {\rm days}}{P_p} \right)^{13/3}
\nonumber \\ & \times &
\left( \frac{P_{\rm dyn}}{2.8\ {\rm hr}} \right)^{10/3}
\left( \frac{T}{5\ {\rm Gyr}} \right). 
\ee
This lower limit on $Q_p$ is well above the canonical value $Q_p \sim 10^5-10^6$,
and thus could confirm the expected large difference in $Q_p$
between forcing in and out of the frequency range of inertial waves.
Outside the inertial range, the 
equilibrium tide calculation gives $Q_p \sim 10^{12}$ 
\citep{2005ApJ...635..688W,1977Icar...30..301G}, so that Earth-size
satellites of hot Jupiters would be immune to destruction by orbital decay
over Gyr timescales.
Note however, that even if $Q_p \sim 10^5-10^6$,
Earth size satellites could exist for longer period orbits $P_p \ga $
a couple months.

% -----------------------------------------------------------

\section{ Heating of the satellites of close-in exoplanets}
\label{sec:heating}

This far we have studied two facets of satellite orbits. First the
variations induced in the satellite orbit by stellar gravity were
reviewed. Then we showed that satellite orbits may be far less susceptible
to orbital decay than previously thought if $Q_p \gg 10^5-10^6$ for
forcing outside the inertial frequency range, as occurs for a satellite
orbit around a synchronized planet. Given that long-lived satellites
may exist around even short-period exoplanets, we now investigate the
consequences of their orbital variation on tidal heating. To motivate
our study, we first review tidal heating in Io.

Based on the significant forced
eccentricity of Io, \citet{1979Sci...203..892P} predicted widespread
surface volcanism, which was soon confirmed by images from Voyager
1 \citep{1979Natur.280..725M}.  The total dissipation rate for
tidal forcing of a homogeneous, incompressible, elastic sphere is
\citep{1978Icar...36..245P,1979Sci...203..892P,2004AJ....128..484W}
\be 
\dot{E} & = & \frac{42}{19} \frac{\pi \rho_s^2 n_s^5 R_s^7 e_{s0}^2}{\mu_s Q_s}
\label{eq:EdotIo}
\ee
where 
$\mu_s \simeq 6.5 \times 10^{11}\ {\rm dyne\ cm^{-2}}$ is the rigidity,
and $Q_s$ is the tidal dissipation
coefficient of Io. For Io, $P_s=1.7{\rm days}$,
$\rho_s = 3.5{\rm g\ cm^{-3}}$, and $R_s=1800\ {\rm km}$.
The eccentricity of Io was determined to be $e_{s0}=0.0043$
\citep{1979Sci...203..892P}, resulting in a dissipation rate for 
Io of 
\be
\dot{E}_{\rm Io} & =& \frac{1.6 \times 10^{21}}{Q_s}\ {\rm erg\ s^{-1}}.
\label{eq:EdotIonum}
\ee
\citet{1979Sci...203..892P} used a fiducial value $Q_s=100$, based on
laboratory studies on rock samples. For $Q_s=100$, eq.\ref{eq:EdotIonum}
gives a heating rate several times the radioactive heating rate estimate
for the Moon \citep{1979Sci...203..892P}. Tidal heating is widely
recognized as an important effect in the thermal structure of Io.

To assess the role of tidal heating for the satellites around hot
Jupiters, we first ask what would happen to an Io-like satellite orbiting
around a Jupiter-like planet, which itself is in 4 day orbit around
a solar-type star. From eq.\ref{eq:deltar} and \ref{eq:deltaphi}, the
perturbations correspond to eccentricity variations $\sim (P_s/P_p)^2 \la
1/25=0.04$. For orbits near the stability limit, the forced eccentricity
is larger than the eccentricity of Io by a factor of $ 0.04/0.0043=9$. The
orbital period of the satellite must be smaller than ${\rm 4\ days}/5=19
{\rm hr}$, shorter than Io's orbital period by a factor of $\simeq
2.1$. Plugging into eq.\ref{eq:EdotIo}, the higher eccentricity and
shorter orbital period combine to increase the heating rate by a factor
$\sim 10^4$. Clearly tidal heating will be important for satellites
around hot Jupiters even for satellite size much smaller than Io.

Tidal heating rates for Earth-like satellites are sensitively dependent
on the tidal $Q_s$. The value $Q_{\rm Earth-Moon} \sim 10$ for the Earth,
as derived from Earth-Moon orbital evolution, is likely due to dissipation
in shallow seas (e.g. \citealt{1970eioh.book.....J}). Heat from such
dissipation depends on the uncertain ocean depth and topography,
and is easily radiated away.  We focus instead on heating of the
interior, especially for early stages to determine if the satellite
is melted. Decay of seismic body waves with periods of $10^3-10^4\
{\rm s}$ in the Earth finds $Q_{\rm Earth\ seismic} \sim 300-500$
(e.g. \citealt{2006JGeo...41..345R}), but extrapolation of this data to
$\sim 10^5-10^6\ {\rm s}$ is uncertain (e.g. \citealt{1977ps...book.....B,
2000MNRAS.311..269E}). For the solid phase, we will use a fiducial value
of $Q_s=100$, assuming moderately increased dissipation when the period
is extrapolated over $10-100$ longer timescales than is measured. As
we will show, the large tidal heating rates for a solid satellite
with $Q_s \sim 100$ imply that satellite is likely melted on a short
timescale. The subsequent tidal heating rate depends on the appropriate
$Q_s$ for the liquid state, which is uncertain. For a completely molten satellite, $Q_s$
may be much larger. For instance, the Preliminary Reference Earth Model,
constructed from seismological data, uses $Q_s \sim 10^5$ for the Earth's
liquid outer core \citep{1981PEPI...25..297D}.

The tidal heating rate in eq.\ref{eq:EdotIo} derived by 
\citet{1979Sci...203..892P} is for the case of a Keplerian satellite
orbit with eccentricity $e_{s0}$. This is only strictly valid for a
satellite orbit unperturbed by external influence. Here we are concerned
with satellite orbits which may be significantly perturbed by stellar
gravity. In other words, we require a tidal heating rate for satellite
orbits given by the solution to the restricted 3-body problem (for small
satellite mass), rather than for a Keplerian orbit. The orbital
variation due to stellar gravity was summarized in section \ref{sec:orbits}.

To discuss the tidal heating rate for satellites perturbed by stellar gravity
we will derive an analytic estimate by plugging
eq.\ref{eq:deltar} and \ref{eq:deltaphi} into 
the formula found in \citet{2004AJ....128..484W}:
\be
\dot{E} & = & \frac{\rho_s h_s R_s^2}{g_s} \int d\Omega\  U \frac{dU'}{dt}.
\label{eq:Edotgeneral}
\ee
This more general formula allows an arbitrary orbit, and is
not specialized to the case of an unperturbed Keplerian orbit.
Here $h_s=(5/2)/(1+19\mu_s/2\rho_s g_s R_s)$, $g_s=GM_s/R_s^2$, $U$ is the
tidal potential in the satellite due to the planet, $U'$ is the tidal potential including the dissipative
lag, and the integral extends over the surface of the satellite. We
perform several operations to simplify eq.\ref{eq:Edotgeneral}. First,
we expand $U$ in spherical harmonics and perform the angular
integration. Next, we expand to leading order in the lag time
$(n_sQ_s)^{-1}$. We assume the satellite is synchronized when evaluating 
the libration term. Lastly, we plug in the small deviations from a circular orbit
found in eq.\ref{eq:deltar} and \ref{eq:deltaphi}.  We find the result
\be
\dot{E} & = & \frac{2817}{160} \frac{GM_p^2h_sR_s^5n_s}{Q_sa_s^6} 
\left( \frac{n_p}{n_s}\right)^4.
\label{eq:tidalheating}
\ee
Note the difference between eq.s \ref{eq:EdotIo} and \ref{eq:tidalheating}. Eq.\ref{eq:EdotIo}
is proportional to $e_{s0}^2$, which is assumed constant for the unperturbed Kepler orbit,
while eq.\ref{eq:EdotIo} is proportional to $(n_p/n_s)^4$, the rms eccentricity squared induced in
the orbit by the stellar gravity. No terms involving $e_{s0}$ appear 
in eq.\ref{eq:tidalheating} as we have perturbed around a Keplerian orbit with $e_{s0}=0$.
Alternatively, a more involved calculation would be to perturb around an orbit with
finite $e_{s0}$, in which case presumably the terms in eq.\ref{eq:EdotIo} and 
eq.\ref{eq:tidalheating} would appear summed together. That is, plugging the Keplerian
result $\delta r=-e_{s0}a_s \cos(n_s t)$ and $\delta \phi_s=2e_{s0} \sin(n_s t)$ into eq.\ref{eq:Edotgeneral}
\footnote{ Equivalently one could use eq.\ref{eq:edottide}, which is simpler in practice. }
would yield eq.\ref{eq:EdotIo}.
%We expect a smooth transition between the eccentricity-dominated case
%given in eq.\ref{eq:EdotIo} and the periodic variation dominated case
%in eq.\ref{eq:tidalheating}. 
Equating these two formulas we find
a critical eccentricity $e_{\rm s, crit} \simeq 1.6 (P_s/P_p)^2 \la 0.06$
above which the term in eq.\ref{eq:EdotIo} would dominate and below which 
the term in eq.\ref{eq:tidalheating} would dominate. The eccentricity 
$e_{s0}$ would depend on the initial conditions as well as subsequent tidal
dissipation in the satellite and planet which would act to decrease $e_{s0}$.
Henceforth we ignore $e_{s0}$ and use the tidal heating rate in eq.\ref{eq:tidalheating}.
Since we have ignored possible finite $e_{s0}$, we may underestimate the heating
rate if $e_{s0}$ is large.

\begin{figure}
\epsscale{1.0}
%\plotone{heatcool_solid_Q=100.ps}
\plotone{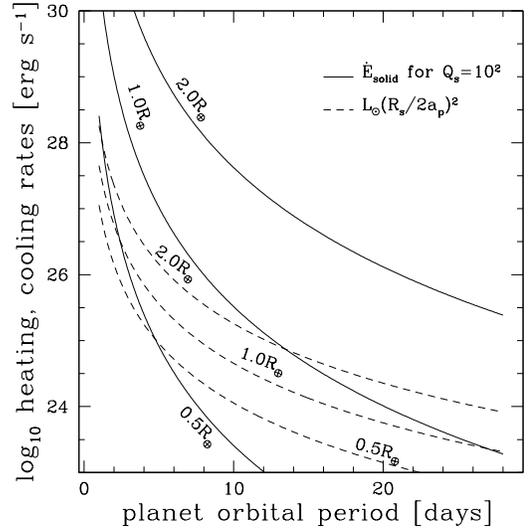}
\caption{ Satellite tidal heating rates (solid lines) as a function of orbital period.
Eq.\ref{eq:Edotsol} was used for the heating rate with
$\rho_s=5\ {\rm g\ cm^{-3}}$, $Q_s=10^2$ and $P_s=P_p/5$. Insolation rates
(dashed lines) for a Sun-like star are shown for comparison.
}
\label{fig:heatcool_solid_Q=100}
\end{figure}

To numerically evaluate eq.\ref{eq:tidalheating}, we again use $\mu_s \simeq 6.5
\times 10^{11}\ {\rm dyne\ cm^{-2}}$ for the rigidity, and scale all
quantities to Earth values. Since the heating rate increases as $P_s$
decreases, we scale expressions to the {\it minimum} heating rate using $P_s=P_p/5$,
finding
\be
\dot{E}_{\rm solid} & = & \frac{3.2 \times 10^{29}\ {\rm erg\ s^{-1}}}{Q_s}
\left( \frac{\rho_s}{5\ {\rm g\ cm^{-3}} } \right)^2
\nonumber \\ & \times & 
\left( \frac{R_s}{R_\oplus} \right)^7 \left( \frac{4\ {\rm days}}{P_p} \right)^5
\left( \frac{P_p/5}{P_s} \right)
\label{eq:Edotsol}
\ee
for a solid satellite and 
\be
\dot{E}_{\rm liquid} & = & \frac{2.7 \times 10^{29}\ {\rm erg\ s^{-1}}}{Q_s}
%\left( \frac{\rho_s}{5\ {\rm g\ cm^{-3}} } \right)^2
\nonumber \\ & \times &
\left( \frac{R_s}{R_\oplus} \right)^5 \left( \frac{4\ {\rm days}}{P_p} \right)^5
\left( \frac{P_p/5}{P_s} \right)
\label{eq:Edotliq}
\ee
for a satellite which has been completely melted ($\mu_s=0$). Eq.\ref{eq:Edotsol} is compared
to insolation in fig.\ref{fig:heatcool_solid_Q=100}. While the prefactors in
eq.\ref{eq:Edotsol} and \ref{eq:Edotliq} are comparable, their scaling
with $R_s$ differs, and the tidal $Q_s$ may be quite different
for molten and solid satellites.
If the tidal $Q_s$ in the liquid phase is far larger than for
the solid phase, then the rate of tidal heating could drop drastically
upon melting.

Would Earth-size satellites around hot Jupiters be melted by tidal
heating?  Eq.\ref{eq:Edotsol} implies a heating rate $\dot{E}_{\rm
solid} \sim 10^{28}\ {\rm erg\ s^{-1}}$ for a 4 day orbital period. For
a latent heat $L=10^{10}\ {\rm erg\ g^{-1}}$ appropriate for iron at
high pressure \citep{1983Icar...54..466S}, melting would occur in a few
hundred years. The tidal heating rate, per gram, scales as $R_s^4$,
so that heating is more intense for larger satellites.  Assuming the
radioactive heating rate of the Earth, $\sim 10^{20}\ {\rm erg\
s^{-1}}$, tidal heating dominates radioactive heating for $P_p \la
150\ {\rm days}$.  Radioactive heating will eventually dominate for
satellites far smaller than an Earth radius. Thermal models of the
young Earth \citep{1983Icar...54..466S} typically show a completely
liquid core for $\dot{E} \ga 10^{20}\ {\rm erg\ s^{-1}}$; the tidal
heating rate is greater by many orders of magnitude. We conclude that
the Earth-size satellites of hot Jupiters, observable by their transits,
would be largely molten, except perhaps for a thin layer near the surface,
cooled by radiation (e.g., \citealt{1979Sci...203..892P}). 
If $Q_s$ drops precipitously for complete melting, 
the temperature could be regulated near the melting point.

\begin{figure}
\epsscale{1.0}
%\plotone{melting_Q=100.ps}
\plotone{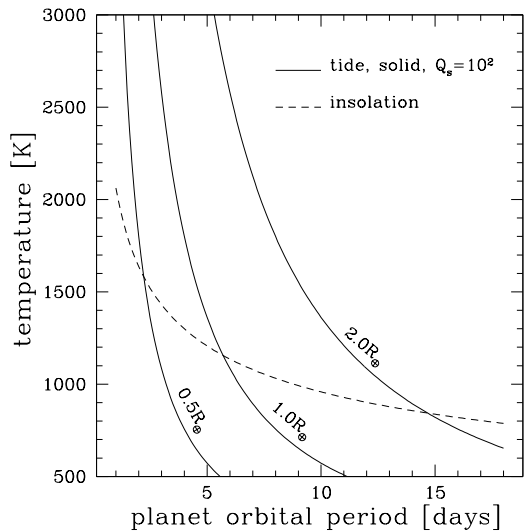}
\caption{ Effective temperature due to tidal heating (solid lines)
or insolation (dashed line). Solid
lines show temperature determined from eq.\ref{eq:Teffsol}, while dashed
lines represent the equilibrium temperature due to insolation, and
reradiation over $4\pi$ steradians.
Tidal heating determines the temperature when the solid line is above
the dashed line.}
\label{fig:melting}
\end{figure}

The effective temperature for the tidal heating energy flux
may be found by equating the heating rate in eq.\ref{eq:Edotsol}
to the blackbody cooling rate $4\pi R_s^2 \sigma T_{\rm eff}^4$, giving
\be
T_{\rm eff} & = & 1800\ {\rm K}
\left( \frac{\rho_s}{5\ {\rm g\ cm^{-3}} } \right)^{1/2}
\left( \frac{R_s}{R_\oplus} \right)^{5/4}
\nonumber \\ & \times &
\left( \frac{4\ {\rm days}}{P_p} \right)^{5/4} \left( \frac{P_p/5}{P_s} \right)^{1/4}
\left( \frac{10^2}{Q_s} \right)^{1/4}.
\label{eq:Teffsol}
\ee
Tidal heating dominates the temperature structure, even at the
surface, when the temperature in eq.\ref{eq:Teffsol} is larger than
the equilibrium temperature implied by insolation.
Figure \ref{fig:melting} compares these temperatures for $Q_s=10^2$,
for a zero-albedo surface with uniform temperature over the entire surface.
Clearly for $Q_s=10^2$, Earth-sized objects are raised above the melting
temperature at zero pressure ($ T_{\rm melt} \la 2000\ {\rm K}$) for either iron or rock compositions
inside a critical orbital period, which we estimate as
\be
P_{\rm p,melt} & = & 3.7\ {\rm days}\
\left( \frac{2000\ {\rm K}}{T_{\rm melt}} \right)^{4/5}
\left( \frac{\rho_s}{5\ {\rm g\ cm^{-3}} } \right)^{2/5}
\nonumber \\ & \times &
\left( \frac{R_s}{R_\oplus} \right)
\left( \frac{10^2}{Q_s} \right)^{1/5}.
\ee
Hence, at the orbital periods of the hot Jupiters (a few days), satellites
Earth size or larger are completely molten for $Q_s=10^2$.

Vigorous tidal heating has consequences for the satellite magnetic field.
Fluid motions in the conducting core driven by tidal heating may generate
magnetic fields through dynamo action.  \citet{1983Icar...54..466S}
derive a magnetic field for the Earth by equating the energy available for
dynamo generation to the Ohmic dissipation rate of the field, which yields
the scaling $B^2 \propto \dot{E}$, where $\dot{E}$ is the tidal heating
rate. As the dissipation rate in eq.\ref{eq:Edotliq} is larger than the
Earth's cooling rate by a factor $\sim 10^9/Q_s$, our results imply that
fields larger than Earth's can be created if $Q_s<10^{9}$.  Intrinsic
magnetic field can decrease the mass loss rate of the atmosphere. Magnetic
interaction between the satellite and planetary magnetosphere may be
of the Jupiter-Ganymede type \citep{1997GeoRL..24.2155K} in which the
standoff distance is determined mainly by magnetic stresses.

Studies of habitability of satellites around gas giants (e.g.,
\citealt{1987AdSpR...7..125R, 1997Natur.385..234W, 2006ApJ...648.1196S})
have invoked tidal heating to maintain plate tectonics, create
subsurface oceans, etc.  They computed satellite tidal heating for
finite eccentricity, assuming it to be a free parameter, or pumped to
large values by interaction with another satellite. The mechanism for
tidal heating in this paper,  periodic eccentricity forced by the star,
is likely unimportant for satellite habitability, since it is only large
for systems very close to the star, where water is already in liquid or
vapor form. Tidal heating is of increasing importance, relative to 
insolation, for host stars of smaller mass since the stellar luminosity 
drops rapidly.

\section{ Energetics }
\label{sec:energetics}

Tidal dissipation in the satellite takes energy out of the satellite
orbit and deposits it in the satellite body as heat. 
%Were there no
%additional source of orbital energy, the satellite's
%orbit would decay into the planet as the orbital energy is decreased.
%For this to occur, the satellite's orbital angular momentum
%must also decrease. Since the satellite's moment of inertia is smaller
%than the orbital moment of inertia by $\sim (R_s/r_s)^2 \la 10^{-4}$,
%the satellite spin rate $\Omega_s$, is quickly synchronized to the orbit
%giving $\Omega_s \simeq n_s$. after which the spin and orbital
%angular momentum are constant. 
In this section we show that the energy
lost from the satellite orbit by heating is exactly balanced by energy
gain from work by the stellar gravity.  The tidal friction acts to create
a lag in the satellite orbital velocity. This lag has the correct phase
to allow energy input to the satellite's orbit at a rate sufficient
to power the tidal dissipation in the satellite. Ultimately, the energy
reservoir for tidal heating of the satellite must then be the star-planet orbit.

Tidal friction acts to exert a velocity dependent acceleration on the reduced
mass of the satellite-planet system (\citealt{2002ApJ...573..829M},
eq.4) with radial and angular components given by 
\be
a_r^{\rm (TF)} & = & - \gamma_0\ \dot{r}_s
\label{eq:arTF}
\\
a_\phi^{\rm (TF)} & = & - \gamma_0\ r_s \left( \dot{\phi}_s-\Omega_s \right).
\label{eq:aphiTF}
\ee
Here $\gamma_0=(9/5)(n_s h_s/Q_s)(m_p/m_s)(R_s/a_s)^5$
is the frictional drag coefficient, which is much smaller than the
orbital mean motion $n_s$. Again we choose a circular background
orbit plus small perturbations as in eq.\ref{eq:rs}, \ref{eq:phis} and \ref{eq:ells}.
Eq.\ref{eq:arTF} and \ref{eq:aphiTF} become
\be
a_r^{\rm (TF)} & = & - \gamma_0\ \delta \dot{r}_s
\label{eq:arTF1}
\\
a_\phi^{\rm (TF)} & = & - \gamma_0\ a_s \left( n_s - \Omega_s + \delta \dot{\phi}_s \right).
\label{eq:aphiTF1}
\ee
Assuming that the satellite spin has already reached the synchronous state, we may set
$\Omega_s=n_s$ in eq.\ref{eq:aphiTF1}.

Adding the tidal friction acceleration in eq.\ref{eq:arTF1} and \ref{eq:aphiTF1} into 
eq.\ref{eq:deltaell} and \ref{eq:ddotdeltar}, and again ignoring the constant radial 
force term, we find equations which have the form of a damped, driven oscillator.
In the limit $\gamma_0 \ll n_s$ we find the solutions
\be
\frac{\delta \dot{r}_s}{a_s} & = & 2 \left( \frac{n_p^2}{n_s} \right)
\sin \left[ 2(n_s-n_p)t + 2\phi_{s0} \right]
\nonumber \\ & + & 
 \frac{35}{6} \gamma_0 \left( \frac{n_p}{n_s} \right)^2
\cos \left[ 2(n_s-n_p)t + 2\phi_{s0} \right]
\label{eq:dotrTF}
\\
\delta \dot{\phi}_s & = &  \frac{11}{4} \left( \frac{n_p^2}{n_s} \right) \cos \left[ 2(n_s-n_p)t +
2\phi_{s0} \right]
\nonumber \\ & - &
 \frac{173}{24} \gamma_0 \left( \frac{n_p}{n_s} \right)^2
\sin \left[ 2(n_s-n_p)t + 2\phi_{s0} \right].
\label{eq:dotphiTF}
\ee
The first terms in eq.\ref{eq:dotrTF} and \ref{eq:dotphiTF} agree with eq.\ref{eq:deltar} and
\ref{eq:deltaphi}. The second terms proportional to $\gamma_0$ describe a lag in the satellite
orbit due to tidal friction.

The energy lost from the orbit is found by computing the work done by the tidal friction force.
Since the tidal friction force already contains the small parameter $\gamma_0 \ll n_s$, we
can ignore terms of order $\gamma_0$ in eq.\ref{eq:dotrTF} and \ref{eq:dotphiTF}.
For reduced mass $\mu_{sp}=M_sM_p/(M_s+M_p) \simeq M_s$, and time averaging so that
$\cos^2, \sin^2 \rightarrow 1/2$, 
we find
\be
\frac{ \dot{E}^{\rm (TF)}_{\rm s,orb} }{\mu_{sp}} & = & 
- \gamma_0 \left( 3 \delta \dot{r}^2_s + a_s^2 \delta \dot{\phi}^2_s \right)
 =  -  \left( \frac{313}{32} \right) \gamma_0\ a_s^2 \left( \frac{n_p^2}{n_s} \right)^2.
\label{eq:edottide}
\ee
This is just a rederivation of eq.\ref{eq:tidalheating}.

The increase in satellite orbital energy due to work done by the stellar gravitational force in 
eq.\ref{eq:deltaell} and \ref{eq:ddotdeltar} is
\be
\frac{ \dot{E}^{\rm (star)}_{\rm s,orb} }{\mu_{sp}} & = &
\frac{3}{2} n_p^2 a_s \left( \delta \dot{r}_s \cos \left[ 2(n_s-n_p)t + 2\phi_{s0} \right]
\nonumber \right. \\ & - &  \left.
 a_s  \delta \dot{\phi}_s \sin \left[ 2(n_s-n_p)t + 2\phi_{s0} \right] \right).
\label{eq:edotstar}
\ee
The terms independent of $\gamma_0$ in eq.\ref{eq:dotrTF} and \ref{eq:dotphiTF}
time average to zero when inserted in eq.\ref{eq:edotstar}. 
Inserting the terms proportional to $\gamma_0$ into eq.\ref{eq:edotstar}
and time averaging we find
\be
\frac{ \dot{E}^{\rm (star)}_{\rm s,orb} }{\mu_{sp}} & = &
\left( \frac{313}{32} \right) \gamma_0\ a_s^2 \left( \frac{n_p^2}{n_s} \right)^2,
\ee
exactly equal and opposite to the energy lost to tidal heating in eq.\ref{eq:edottide}.

We have found that energy lost from the orbit is replenished by work done
on the orbit by the stellar tidal force. Hence the satellite orbit is
stable on long timescales to tidal effects arising from dissipation
in the satellite. The energy reservoir powering the tidal dissipation in
the satellite must then come from the star-planet orbit. For star-planet
orbital energy
\be
E_{\rm \star p} & = & - \frac{GM_\star M_p}{2a_p}
\nonumber \\ & = & 
- 1.8 \times 10^{44}\ {\rm erg}\ \left( \frac{M_\star}{M_\odot} \right)^{2/3}
\left( \frac{M_p}{M_{\rm Jup}} \right) \left( \frac{4\ {\rm days}}{P_p} \right)^{2/3},
\ee
and dissipation in a liquid satellite (eq.\ref{eq:Edotliq}), the star-planet orbit
is immune to orbital decay on a timescale $T$ if
\be
Q_s & \geq & 480\ \left( \frac{M_\star}{M_\odot} \right)^{-2/3}
\left( \frac{M_p}{M_{\rm Jup}} \right)^{-1} \left( \frac{R_s}{R_\oplus} \right)^5
\nonumber \\ & \times &
\left( \frac{4\ {\rm days}}{P_p} \right)^{13/3} \left( \frac{P_p/5}{P_s} \right)
\left( \frac{T}{10^{10}\ {\rm yr}} \right).
\label{eq:Qslim}
\ee
This critical value of $Q_s$ for orbital decay is comparable to
the value expected for a solid Earth-like satellite (see section
\ref{sec:heating}). In the more likely case that the satellite has been
melted by tidal heating and $Q_s \gg 10^2$, the star-planet orbit is
immune to decay.

\section{ Evaporative mass loss }
\label{sec:massloss}

The large tidal heating rates in eq.\ref{eq:Edotsol} and \ref{eq:Edotliq},
large insolation from the parent star, and low escape speed raise the possibility
of enhanced rates of mass loss from the satellites studied in this
paper. The rates of mass loss are difficult to predict, and may depend
on many factors such as stellar wind or magnetospheric erosion of the
atmosphere, and the flux of vapor from the surface to the exosphere. 
Here we discuss one possible scenario, of thermal evaporation
from the solid surface of a satellite with a negligible atmosphere
(such as the Moon or Mercury).

\citet{1985Icar...64..285C} and \citet{1987E&PSL..82..207F} discussed
evaporative mass loss from Mercury in the protoplanetary nebula as a means
of understanding Mercury's high mean density. The high temperatures in the
nebula allow enhanced rates of evaporation of low(er) density silicates
from the solid surface into gaseous form. If this vapor is removed from
the atmosphere, the mass fraction of iron in the body, and hence the mean
density, will increase.  Due to the exponential temperature dependence,
this process is only effective at high temperatures.  We find that for
surface temperatures due to tidal heating, or the equilibrium temperature
due to insolation, that significant mass loss can occur. We will follow
the simple estimate in \citet{1985Icar...64..285C} to estimate the
importance of this process.

The number flux for thermal evaporation from a surface, in ${\rm cm^{-2}\ s^{-1}}$, is given by
\be
\phi & = & \frac{P_{\rm eq}(T)}{\sqrt{2\pi m k_b T}},
\label{eq:rate}
\ee
where $P_{\rm eq}(T)=P_010^{-T_0/T}$ is the equilibrium vapor pressure above the surface,
$m$ is the mean molecular weight (in grams) of the products, $T$ is the temperature, and $P_0$ and $T_0$
describe the equilibrium pressure. \citet{1985Icar...64..285C} considers
an illustrative example of ${\rm MgSiO_3(liq) = Mg(gas) + SiO(gas) + O_2(gas)}$
(see \citet{1987E&PSL..82..207F} for more detailed calculation of silicate magmas of 
chondritic composition), for which $P_0=10^{13.176}\ {\rm dyne\ cm^{-2}}$,
$T_0=24605\ {\rm K}$, and $m=5.5\times 10^{-23}\ {\rm g}$.
If the surface temperature is near the equilibrium temperature
for a zero albedo surface (neglecting tidal heating), radiating with the same efficiency over $4\pi$
\footnote{ We note that if the atmosphere is thin (surface pressure $\ll {\rm 1\ bar}$),
the day-side temperature may be larger
by a factor of $2^{1/2}$. This will cause the critical $a_p$ for large mass loss to 
increase by a factor of 2, and the critical planet orbital period to increase by a factor
$2^{3/2}$. Hence a critical orbital period inside of which large mass loss occurs depends
sensitively on the $2^{1/2}$, since it occurs in an exponent. Our assumption of equal day-night
temperatures is conservative.},
the equilibrium temperature versus orbital radius for a solar type star is $T=5777\ {\rm K}
(R_\odot/2a_p)^{1/2}=1300\ {\rm K} (10R_\odot/a_p)^{1/2}$. As seen in figure \ref{fig:melting},
this temperature including only insolation, and ignoring tidal heating, may underestimate
the satellite surface temperature close to the star. Approximating the 
exponential as a power law (valid for small changes in $a_p$), we find a mass flux
\be
m\phi & \simeq & 1.3 \times 10^{-11}\ {\rm g\ cm^{-2}\ s^{-1}} \left( \frac{10R_\odot}{a_p}
\right)^{21.5}
\ee
from the surface. If the vapor is lost from the atmosphere, 
this translates into a mass loss rate
\be
\dot{M_s} & = & 3.3 \times 10^{-4} {\rm M_\oplus\ Gyr^{-1}} \left( \frac{10R_\odot}{a_p}
\right)^{21.5} \left( \frac{R_s}{R_\oplus} \right)^2.
\label{eq:Mdot}
\ee
In $5\ {\rm Gyr}$, an Earth mass can be lost for $a_p \la 7.4 R_\odot$,
where $T \ga 1500\ {\rm K}$.  Hence satellites originally of Earth-size,
and hence detectable by their transit signature with Kepler, may become
undetectable due to erosion. Eq.\ref{eq:Mdot} implies the rate of 
decrease of the radius is independent of radius. Erosion does not
slow for small satellites.

Our estimates give an upper limit to the erosion rate, and show that large
mass loss can occur over the lifetime of the satellite. The exponential
dependence on surface temperature suggests a critical orbital radius (of
planet around the star) inside of which satellites can be significantly
eroded. Our conservative estimate finds the critical orbital period to
be around 2.4 days (6.6 days using the day-side equilibrium temperature).

\section{ Conclusions }
\label{sec:conclusions}

In this paper we have studied the orbits, tidal heating and evaporative
mass loss rates from Earth-sized satellites orbiting hot Jupiters.
After reviewing the perturbation to satellite orbits due to the stellar
gravity, we derived four main results.  First, we have shown that orbital
decay due to tides raised in the planet by the satellite may be much
less efficient that previously thought since gas giants are likely very
weakly dissipative at the forcing frequencies of interest.  Consequently,
even Earth-size satellites may be stable around hot Jupiters over Gyr
timescales. Second, large tidal dissipation rates are induced in the
satellite due to its forced orbital variations, likely melting all but
perhaps a thin surface layer. Third, we found that the satellite orbit
does not evolve secularly due to tidal dissipation in the satellite,
as the stellar gravity does work at a rate to keep the orbital energy
constant. Lastly, the estimated upper limit to mass loss, valid for
sufficiently thin atmospheres, indicates a critical orbital period around
a Sun-like star (2.4 days for a thick atmosphere, 6.6 days for a thin
atmosphere) for Earth-like satellites orbiting a hot Jupiter. Inside
this critical period significant erosion can occur, reducing satellites
to a size undetectable by upcoming transit observations with the Kepler
satellite. Outside the critical period, evaporative mass loss will become
negligible so that detectable Earth size planets can survive.

%--------------------------------------------------------------------------------
\acknowledgements

We thank Ken Seidelmann for interesting discussions on lunar theory.
TAC was supported by a NASA Graduate Student Researchers Program
Fellowship through the Langley Research Center and a Virginia Space
Grant Consortium Graduate Research Fellowship. RM received support 
through a Virginia Space Grant Consortium Undergraduate Scholarship.
PA is an Alfred P. Sloan Fellow, and also acknowledges support from the
University of Virginia Fund for Excellence in Science and Technology.
REJ acknowledges support from NASA's Planetary Atmospheres Program.
We thank the anonymous referee for constructive comments which improved
this paper significantly.

%--------------------------------------------------------------------------------

\end{document}